\documentclass{article}
\usepackage[T1]{fontenc} 
\usepackage[utf8]{inputenc} 
\usepackage{ismir,amsmath,cite,url,mathrsfs}
\usepackage{graphicx}
\usepackage{color}
\usepackage{algorithm}
\usepackage{lineno}
\usepackage{booktabs}
\usepackage{multirow}
\usepackage{amsfonts}
\usepackage{arydshln}

\makeatletter
\newcounter{phase}[algorithm]
\newlength{\phaserulewidth}

\makeatother

\title{Predicting performance difficulty from \\ piano sheet music images}

\multauthor
{Pedro Ramoneda$^1$ \hspace{1cm} Jose J. Valero-Mas$^1$ } { \bfseries{Dasaem Jeong$^2$} \hspace{1cm} \bfseries{Xavier Serra$^1$ }\\
$^1$ Music Technology Group, Universitat Pompeu Fabra, Barcelona\\
{\tt\small \{pedro.ramoneda, josejavier.valero, xavier.serra\}@upf.edu}\\
$^2$ MALer Lab, Sogang University, Seoul\\
{\tt\small dasaemj@sogang.ac.kr}
}

\def\authorname{P. Ramoneda, J. J. Valero-Mas, D. Jeong and X. Serra}

\begin{document}

\maketitle
\begin{abstract}
Estimating the performance difficulty of a musical score is crucial in music education for adequately designing the learning curriculum of the students. Although the Music Information Retrieval community has recently shown interest in this task, existing approaches mainly use machine-readable scores, leaving the broader case of sheet music images unaddressed. Based on previous works involving sheet music images, we use a mid-level representation, bootleg score, describing notehead positions relative to staff lines coupled with a transformer model. This architecture is adapted to our task by introducing an encoding scheme that reduces the encoded sequence length to one-eighth of the original size. In terms of evaluation, we consider five datasets---more than 7500 scores with up to 9 difficulty levels---, two of them particularly compiled for this work. The results obtained when pretraining the scheme on the IMSLP corpus and fine-tuning it on the considered datasets prove the proposal's validity, achieving the best-performing model with a balanced accuracy of 40.34\% and a mean square error of 1.33. Finally, we provide access to our code, data, and models for transparency and reproducibility.
\end{abstract}

\section{Introduction}
\label{sec:introduction}

Estimating the difficulty of a piece is crucial for music education, as it enables the effective structuring of music collections to attend to the student's needs. This has led to a growing research interest~\cite{sebastien2012score,chiu2012study,nakamura2018statistical,anonym}, as well as the development of automatic systems for exploring difficulties by major industry players such as Muse Group~\cite{musescore,ultimate} and Yousician~\cite{kaipainen2017system}.

\begin{figure}[t]
  \centering
  \includegraphics[width=0.9\linewidth]{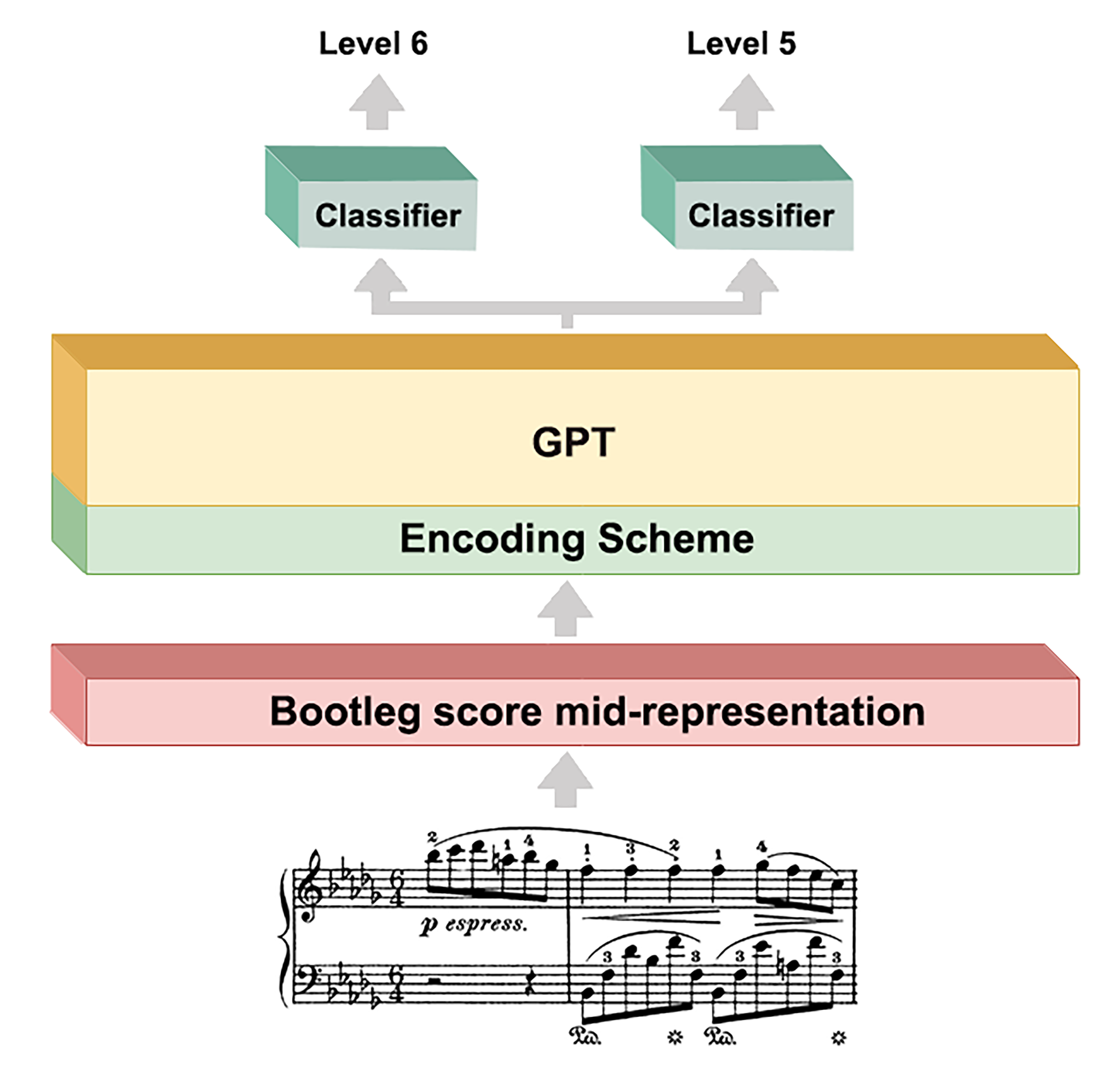}
  \caption{We consider the bootleg score mid-representation with a multi-task GPT-based recognition framework to predict the performance difficulty associated to a piano score directly from sheet images from multiple annotated collections with varied difficulty levels.}
  \label{fig:example}
\end{figure}

Previous research on predicting piano difficulty has primarily focused on symbolic machine-readable scores~\cite{sebastien2012score,chiu2012study,nakamura2014merged,nakamura2015automatic,ramoneda2022,anonym}. Early  studies explored feature engineering descriptors~\cite{sebastien2012score, chiu2012study} and the relationship between piano fingering and difficulty~\cite{nakamura2014merged,nakamura2015automatic,ramoneda2022}. A recent study~\cite{anonym} used stacked recurrent neural networks and context attention for difficulty classification on machine-readable scores, employing embeddings from automatic piano fingering, piano expressive generation~\cite{jeong2019virtuosonet}, and score information. This study found that modeling the score difficulty classification task as an ordinal regression problem~\cite{cheng2008neural} was advantageous, and using entire pieces for training, rather than fragments, was essential to avoid degraded performance.

Although symbolic machine-readable scores offer more interpretability~\cite{ramoneda2022}, with all the music information completely accessible, their limited availability compared to sheet music images restricts the practical use of difficulty prediction tools for librarians, teachers, and students. Focusing on sheet music image analysis expands the range of available music, has the potential to preserve the cultural heritage of symbolic-untranscribed scores, and addresses the lack of diversity in Western classical piano curricula. By analyzing image-based sheet music, we aim to create technology for highlighting historically underrepresented communities like female composers~\cite{bennett2018creating, halstead2017woman} and promoting diversity in piano education. This promotion is crucial since the piano teaching repertoire has remained mostly unchanged for decades~\cite{cutietta2007content}, containing around 3,300 pieces~\cite{magrath1995pianists}, while projects such as IMSLP house remarkably larger databases.

One of the main challenges in working with sheet music is attaining a symbolic music-based representation for direct analysis. Although Optical Music Recognition (OMR) literature has considerably improved in creating such representations over the past 30 years, it remains an unsolved  task~\cite{calvo2020understanding}. Bootleg score~\cite{imslp2019tsai} is an alternative to symbolic scores obtained with OMR. This mid-level symbolic representation keeps the most relevant primitives of the music content in a music sheet, which has shown remarkable success in several tasks~\cite{large2022tsai,alignment2019tsai,ismir2021sheet,tismir2022tsai}, especially in classification, such as piano composer classification~\cite{ismir2020tsai,ismir2021tsai,large2022tsai} or instrument recognition~\cite{ismir2021instrument}.

We build on this literature, employing the GPT model~\cite{radford2019language} and bootleg score in our analysis. More precisely, we consider the approach by Tsai et al.~\cite{imslp2019tsai}, in which a GPT model pretrained on the IMSLP piano collection is finetuned for specific recognition tasks. With adequate adaptations, we hypothesize that this framework may also succeed in estimating performance difficulty on music sheet images.

As aforementioned, difficulty estimation benefits from the use of entire music pieces rather than excerpts to obtain adequate success rates. However, processing large sequence stands as a remarkable challenge in music processing, especially when addressing bootleg representations due its considerable verbosity. While some recent mechanisms address this issue in general learning frameworks (\textit{e.g.}, Flash Attention~\cite{dao2022flashattention}), we extend the original proposal by Tsai et al.~\cite{imslp2019tsai} with a multi-hot optimization target for GPT pretraining, and replace the categorical encoding with causal convolutional or feedforward projection layers to enhance performance and reduce costs.

Moreover, addressing data scarcity is crucial for promoting and establishing this task within the Music Information Retrieval community. As of now, the \emph{Mikrokosmos-difficulty}~(MK)~\cite{ramoneda2022} and \emph{Can I Play It?}~(CIPI)~\cite{anonym} symbolic datasets stand for the only available annotated collections, out of which music sheet images can be obtained by engraving mechanisms. To enhance data availability and encourage further research, we have collected additional datasets from existing collections, namely \emph{Pianostreet-difficulty} (PS), \emph{Freescore-difficulty} (FS), and black female composers collection Hidden Voices (HV). This results in more than 7500 music pieces, spanning up to 9 difficulty levels and each annotated with a difficulty classification system. 
Although difficulty prediction contains a subjective element, global trends may emerge when examining multiple difficulty classification systems simultaneously. To our knowledge, no previous research has explored this aspect. Consequently, we propose a multitask approach to training simultaneously on CIPI, PS, and FS datasets. 
Finally, we also analyze the generalization of our proposed methodologies with the MK and HV benchmark datasets.

Considering all above, our precise contributions are: (i) we adopt the previous bootleg-representation literature~\cite{ismir2020tsai,ismir2021tsai}, pretraining a GPT model on IMSLP and finetuning it for our task, adapting the encoding scheme accordingly, as presented in Figure~\ref{fig:example}; (ii) we evaluate our proposal using a novel sheet music image collection of five datasets with more than 7,500 pieces with difficulty levels ranging up to 9; (iii) we propose a multi-task strategy for combining multiple difficulty classification systems from the datasets; (iv) we conduct extensive experiments to assess the proposed methodologies, including a zero-shot scenario for testing generalization and comparisons with previous proposals on the CIPI dataset; and (v) to promote the task, code, and models~\footnote{\url{https://github.com/PRamoneda/pdf-difficulty}}, and datasets~\footnote{\url{https://zenodo.com/record/8126801}} are publicly available.

\section{Music sheet image datasets}
\label{sec:dataset}

Due to the relative recentness of the field,  the lack of annotated corpora has severely constrained the performance difficulty assessment. The earliest data assortments may be found in the works by Sebastian et al.~\cite{sebastien2012score} and Chiu et al.~\cite{chiu2012study}, which respectively collected 50 and 300 MIDI scores from different score repositories. However, these datasets were never publicly released.

To our best knowledge, the \emph{Mikrokosmos difficulty} (MK) set by Ramoneda et al.~\cite{ramoneda2022}, which comprises 147 piano pieces by Béla Bartók in a symbolic format graded by the actual composer, represents the first publicly available collection for the task at hand. More recently, the authors introduced the \emph{Can I Play It?} (CIPI) dataset~\cite{anonym}, a collection of 652 piano works in different symbolic formats annotated after 9 different difficulty levels. Note that, while sheet music scores can be obtained by resorting to engraving mechanisms, the insights obtained may not apply to real-world scenarios.

\begin{table}[h!]
\centering
\addtolength{\tabcolsep}{-0.2em}
\resizebox{\columnwidth}{!}{%
\begin{tabular}{lccccc}
\toprule[1pt]
\textbf{Dataset} & \textbf{Pieces} & \textbf{Classes} & \textbf{AIR} & \textbf{Noteheads} & \textbf{Composers}\\
\cmidrule(lr){1-6}
MK~\cite{ramoneda2022} & 147 & 147 & .78 & 49.2k & 1\\
CIPI~\cite{anonym} & 652 & 9 & .33 & 1.1M & 29\\
\hdashline
PS & 2816 & 9 & .24 & 7.2M & 92\\
FS & 4193 & 5 & .37 & 5.8M & 747\\
HV & 17 & 4 & 1 & 21.5k & 10\\
\bottomrule[1pt]
\end{tabular}
}
\caption{Description of existing collections for performance difficulty estimation based on the number of pieces, classes, average imbalance ratio (AIR), noteheads, and composers. The dashed line differentiates the datasets based on symbolic (above) and image (below) sheet music.}
\label{tab:dataset}
\end{table}

To address this limitation, we compiled a set of real sheet music images of piano works together with their performance difficulty annotations from different music education and score-sharing platforms on the Internet. More precisely, we arranged three different collections attending to the source: (i) the \emph{Pianostreet-difficulty} (PS) set retrieved from~\cite{ps_web} that depicts 2,816 works with 9 difficulty levels annotated by the Pianostreet team; (ii) the \emph{Freescores-difficulty} (FS) assortment from~\cite{fs_web} that contains 4,193 pieces with 5 difficulty levels comprising a variety of compositions and annotations by the users of the platform; and (iii) the \emph{Hidden Voices} (HV) collection~\cite{hidden,walkerbook}, a set of 17 pieces by black female composers annotated with 4-level difficulty labels by musicologists of the Colorado Boulder Music Department.


Table~\ref{tab:dataset} summarizes the main characteristics of commented publicly-available collections. The \textit{average imbalance ratio} (AIR), measured as the mean of the individual ratios between each difficulty class and the majority label in each collection, is also provided for reference purposes.

\section{Methodology}
\label{sec:methods} 
Based on its success when addressing classification tasks from sheet music images~\cite{ismir2020tsai,ismir2021instrument}, our proposal considers the use of the so-called bootleg score representation coupled with a GPT-based recognition model to estimate the performance difficulty of a piece.

Introduced by~\cite{imslp2019tsai}, bootleg scores stand as a simple---yet effective---representation to encode the content of a sheet music image for certain recognition tasks. Formally, a bootleg score is a binary matrix of length $w$ and $h=62$ vertical positions---\textit{i.e.}, $\mathcal{X}\in\left\{0,1\right\}^{w\times 62}$---that respectively denote the temporal and pitch dimensions. Note that the $w$ value represents the number of note heads detected by the bootleg extraction process. Our work resorts to this representation, being the use of alternative codifications posed as a future line to address.

The GPT recognition framework undergoes an unsupervised pretraining step on the IMSLP piano collection, which was originally used by \cite{imslp2019tsai}. Eventually, considering a set of labeled data $\mathcal{T}\subset \mathcal{X} \times \mathcal{C}$ where $\mathcal{C}=\left\{c_{1},\ldots, c_{|\mathcal{C}|}\right\}$ denotes the possible difficulty levels, the model is finetuned to retrieve the recognition function $\hat{f}:\mathcal{X}\rightarrow\mathcal{C}$ that relates a bootleg representation to a particular difficulty level. Based on previous work addressing this task~\cite{anonym}, we consider an ordinal classification framework~\cite{cheng2008neural} as the difficulty grading scales naturally fit this formulation.

Despite being capable of addressing the task, the framework was noticeably affected by two factors: (i) the excessive length of the input sequences when pretraining the model; and (ii) the inconsistent definition of difficulty levels among corpora. Consequently, we introduce two mechanisms specifically devised to address these limitations.

\subsection{Sequence length in pretraining} 
\label{method:schemas}

One of the main drawbacks related to bootleg representations is their verbosity, as it depicts $h=62$ elements per frame. To address this issue, Tsai et al.~\cite{ismir2020tsai} proposed subdividing each column into groups of 8 elements and encoding each according to a vocabulary of $|\sigma|=2^{8}$ elements. In this regard, the initial bootleg score $x\in\left\{0,1\right\}^{w\times 62}$ is mapped to a novel space defined as $\Sigma^{w\times 8}$. This representation is then flattened to undergo a categorical embedding process that maps it to a feature-based space denoted as $\mathbb{R}^{8w\times 768}$, which is eventually used for pretraining the GPT model with 768-dim hidden states. Note that this process reduces the vocabulary size and remarkably increases the sequence length.

To address this issue, we propose substituting this tokenization process with an embedding layer that directly maps the bootleg score into a suitable representation, avoiding the extension of the initial length of the sequence. In this sense, the initial bootleg representation $x\in\left\{0,1\right\}^{w\times 62}$ is mapped to a space defined as $\mathbb{R}^{w\times 768}$ that serves as input to the GPT model with a fraction of the length of the encoding used by Tsai et al.~\cite{ismir2020tsai}. Besides reducing the length of the sequences to process, we hypothesize that such an embedding may benefit the recognition model as a suitable representation is inferred for the task. In this regard, our experiments will compare two types of embedding approaches---more precisely, a fully-connected layer and a convolutional one, respectively denoted as FC and CNN---to quantitatively assess this claim.

\begin{figure}[h!]
  \centering
  \includegraphics[width=0.9\linewidth]{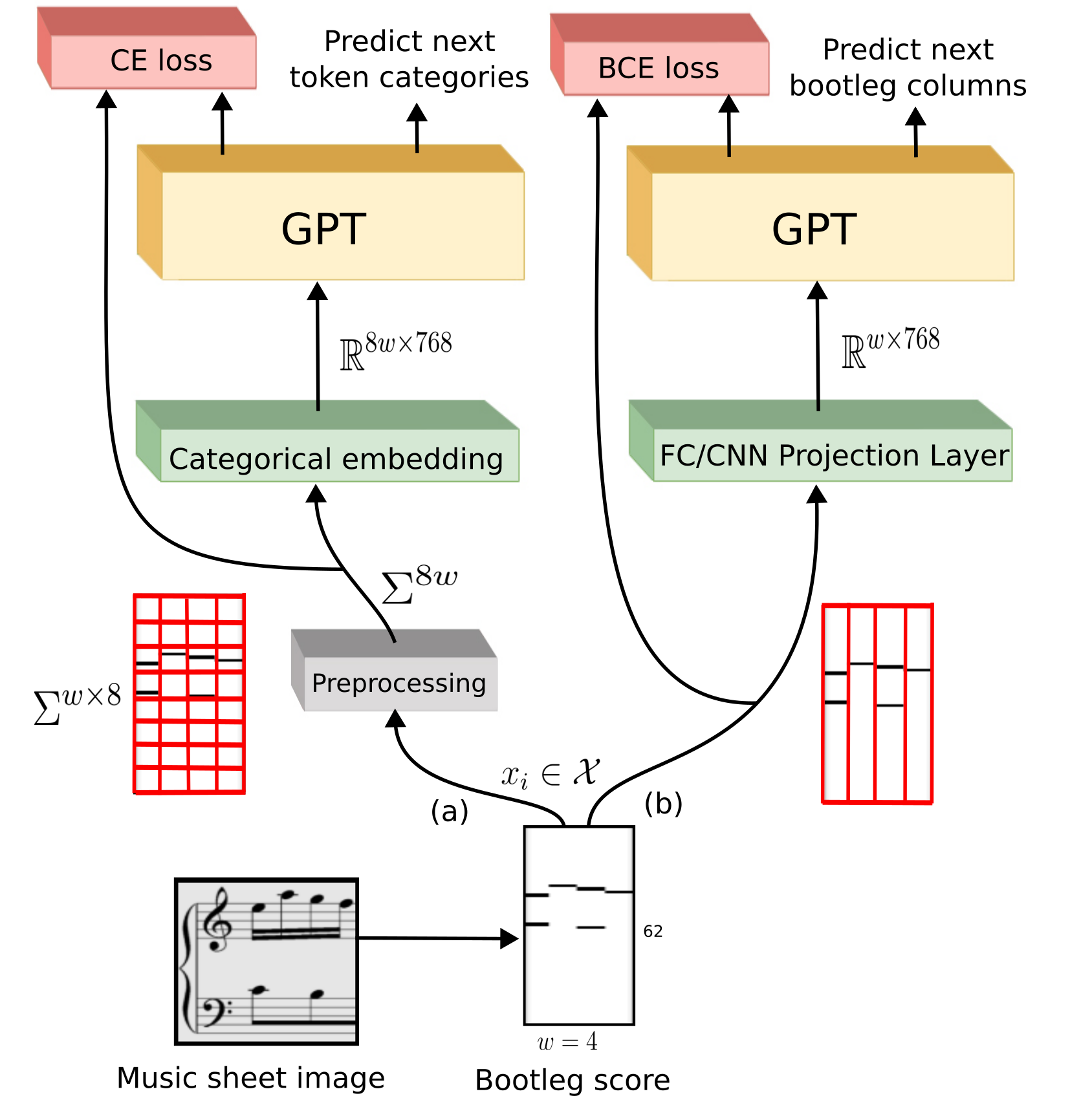}
  \caption{Comparison between the proposal by Tsai et al.~\cite{ismir2020tsai}---denoted as (a)---and the presented proposal---highlighted as (b)---for a case of toy example with a duration of $w=4$.}
  \label{fig:pretraining}
\end{figure}

Figure~\ref{fig:pretraining} graphically describes the approach by Tsai et al.~\cite{ismir2020tsai} and the presented proposal. In opposition to the reference work, the proposal considers multi-hot encoding instead of discrete categorical index as the output of the GPT recognition framework, by using binary cross-entropy loss instead of negative log-likelihood loss.


\subsection{Multi-task learning of multiple difficulty classification systems} 
\label{method:multi}

\begin{figure}[h!]
  \centering
  \includegraphics[width=1\linewidth]{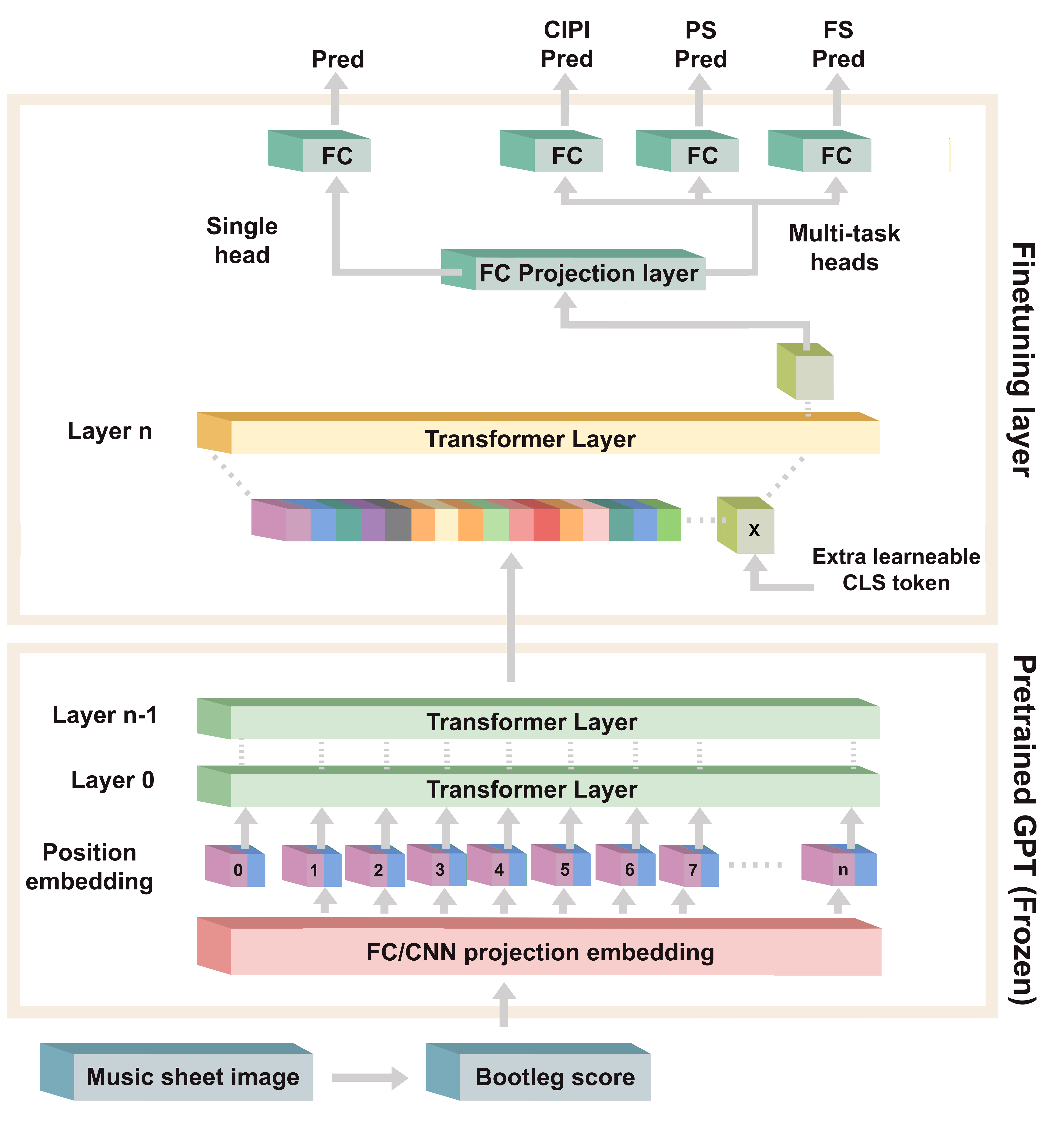}
  \caption{Graphical description of the downstream architecture depicting the classification heads for the multi-task proposals as well as the single-head case of the reference work.}
  \label{fig:down}
\end{figure}

The pretrained GPT model can be simply finetuned for a performance difficulty classification task by adding a projection layer and a learnable classification token, as depicted in Figure~\ref{fig:down}. However, the actual definition of the performance difficulty of a piece is a highly subjective problem that may bias---and, hence, remarkably hinder---the goodness of a recognition model. In this regard, we hypothesize that using a multi-task approach that attends different definitions of difficulty---\textit{i.e.}, a labeled assortment of data from multiple annotators---may benefit the generalization capabilities of the approach.

In this regard, we modify the reference architecture for the downstream task to include an additional classification layer for each training collection. While simple, such a proposal is expected to improve the overall recognition performance given the wider variety of data provided during the training process. Figure~\ref{fig:down} graphically describes this proposal.

Finally, no pre-processing is done in relation to the label distribution of the corpora to avoid inducing any type of bias. In this regard, the sampling protocol of the model has been forced to maintain its original distributions. 

\newcommand{\tj}{$\text{GPT}_\texttt{EMB}$}
\newcommand{\cnn}{$\text{GPT}_\texttt{CNN}$}
\newcommand{\fc}{$\text{GPT}_\texttt{FC}$}
\newcommand{\multifc}{$\text{GPT}^\texttt{multi}_\texttt{FC}$}
\newcommand{\multicnn}{$\text{GPT}^\texttt{multi}_\texttt{CNN}$}

\section{Experimental setup}
\label{exp:setup}

\subsection{Data collections and assessment metrics}

To validate the proposal, we have considered the five publicly-available data collections presented in Section~\ref{sec:dataset}, \textit{i.e.}, \emph{Mikrokosmos difficulty} (MK)~\cite{ramoneda2022}, \emph{Can I Play It?} (CIPI)~\cite{anonym}, \emph{Pianostreet-difficulty} (PS)~\cite{ps_web}, \emph{Freescores-difficulty} (FS)~\cite{fs_web}, and \emph{Hidden Voices} (HV)~\cite{hidden,walkerbook}. While MK and CIPI exclusively comprise symbolic scores, we engraved them into music sheets and included them due to the commented scarcity of annotated data.

We considered a 5-fold cross-validation scheme with a data partitioning of 60\% for the finetuning phase after the pretraining stage with IMSLP together with two equal-size splits of the remaining data for validation and testing. Note that, since MK and HV are exclusively used for benchmark purposes, no partitioning is applied to them.

In terms of performance evaluation, we resort to two assessment criteria typically used in ordinal classification~\cite{gaudette2009evaluation}: \textit{accuracy within $n$} ($\mbox{Acc}_{n}$) and \textit{mean squared error} (MSE). To adequately described them, let $\mathcal{S} \subset\mathcal{X}\times{C}$ denote a set of test data and let $\mathcal{S}_{c}=\left\{(x_{i},y_{i})\in\mathcal{S}:y_{i} = c\right\}$ with $1\leq i\leq |\mathcal{S}|$ be the subset of elements in $\mathcal{S}$ with class $c$.

Based on this, $\mbox{Acc}_{n}$ is defined as:
\begin{equation}
    \mbox{Acc}_{n} = \frac{1}{|\mathcal{C}|}\sum_{\forall c\in\mathcal{C}}\frac{\left|\left\{y\in\mathcal{S}_{c}: \left|\hat{f}(x) - c\right|\leq n\right\}\right|}{\left|\mathcal{S}_{c}\right|}
\end{equation}
\noindent where $\hat{f}(\cdot)$ represents the trained recognition model and $n\in\mathbb{N}_{0}$ denotes the tolerance or class-boundary relaxation that allows for errors in adjacent labels. In our experiments we consider the values of $n=0$ (no tolerance) and $n=1$ (smallest adjacency tolerance), respectively denoted as $\mbox{Acc}_{0}$  and $\mbox{Acc}_{1}$ in the rest of the work.

Regarding MSE, this figure of merit is defined as: 
\begin{equation}
    \mbox{MSE} = \frac{1}{|\mathcal{C}|}\sum_{\forall c\in\mathcal{C}}\frac{\sum_{\forall x\in \mathcal{S}_{c}}\left(\hat{f}(x)-c\right)^{2}}{\left|\mathcal{S}_{c}\right|}
\end{equation}

Finally, note that all these metrics are macro-averaged to account for the unbalanced nature of the data collections used in the work.


\subsection{Training procedure}
As commented, the recognition model undergoes an initial pretraining stage considering the IMSLP corpus. During this stage, the model considers sequences of 256 tokens, each with a binary cross-entropy as a loss function. To speed up this process, the Flash Attention framework by~\cite{dao2022flashattention} is also considered. For comparative purposes, all other parameters remain unaltered from the reference works~\cite{ismir2020tsai}.

After that, the model is finetuned on the downstream difficulty estimation task, considering an Adam optimizer~\cite{kingma2014adam} with a learning rate of $10^{-5}$ and early stopping based on the Acc$_0$ and MSE metrics on the validation set. Moreover, a balanced sampler is considered to tackle the issue of unbalanced data collections. Ordinal Loss~\cite{cheng2008neural} is applied to train the difficulty prediction as an ordinal classification problem, while no loss weighting considered in the multi-task framework.  For regularization and stable training, gradient clipping is set to $10^{-4}$, with a batch size of 64 and L2 regularization. This optimization process is carried out exclusively on the last layer of the model, resorting the remaining parts to the weights obtained during the pretraining phase of the procedure. 

Note that while these processes may be further studied to account for the optimal solution that retrieves the best-performing results, such a study is out of the scope of the work and is left as future work to address.

\section{Experiments and results}
\label{sec:experiments}

This section presents the results obtained with the introduced experimental scheme. To adequately provide insights about the task, the section provides a series of individual experiments devoted to analyzing one aspect of the proposal: Section~\ref{exp:scheme} analyzes the influence of the encoding scheme; Section~\ref{exp:multi} evaluates  the influence of the multitask architecture; Section~\ref{exp:zero} delves on the ranking generalization in a zero-shot scenario; finally, Section~\ref{exp:cipi} compares the attainable results when addressing the task from the symbolic versus the sheet-image domains.

\subsection{Encoding schemes experiment}
\label{exp:scheme}

This first experiment compares the performance of the two encoding schemes presented in Section~\ref{method:schemas}, \textit{i.e.}, \fc{} and \cnn{}. Table~\ref{tab:representation} presents the results obtained for the CIPI, FS, and PS collections for the three figures of merit considered.



\begin{table}[h!]
\centering
\resizebox{.9\columnwidth}{!}{%
\begin{tabular}{llccc}
\toprule[1pt]
\multicolumn{2}{l}{Encoding}  & Acc$_{0}$ (\%)      & Acc$_{1}$ (\%) & MSE      \\ \hline
\multicolumn{2}{l}{\textit{Can I Play it?}}\\
& $\text{GPT}_\texttt{FC}$  & 34.3(6.1) & 78.1(4.6) & 1.6(0.3) \\
& $\text{GPT}_\texttt{CNN}$  & \textbf{36.2(8.2)} & \textbf{81.7(1.5)} & \textbf{1.4(0.1)} \\ 
\hdashline
\multicolumn{2}{l}{\textit{PianoStreet}}\\
& $\text{GPT}_\texttt{FC}$   & 30.9(3.8) & 71.1(9.6) & 2.1(0.4) \\
& $\text{GPT}_\texttt{CNN}$  & \textbf{31.8(1.6)} & \textbf{78.8(1.8)} & \textbf{1.9(0.1)} \\
\hdashline
\multicolumn{2}{l}{\textit{FreeScores}}\\
& $\text{GPT}_\texttt{FC}$   & 46.6(1.9)           & \textbf{92.5(1.0)} & \textbf{0.8(0.1)} \\
& $\text{GPT}_\texttt{CNN}$  & \textbf{47.3(3.4)}           & 92.4(0.6) & 0.8(0.1) \\ 
\bottomrule[1pt]
\end{tabular}
}
\caption{Results of comparing the encoding schemes \fc{} and \cnn{}. Bold values highlight the best results per collection and metric.}
\label{tab:representation}
\end{table}

As it may be observed, the \cnn{} experiment outperformed the \fc{} experiment in most evaluation metrics across the three datasets. More precisely, the \cnn{} consistently achieved the best performance in the Acc$_{0}$ metric for all data collections, showing an average improvement of $1\%$ concerning the \cnn{} case. This trend remains for the rest of the figures of merit except for the case in the FS assortment, in which the results of the FC-based model outperform those of the CNN case. 

Nevertheless, attending to the high standard deviations, the performance results of the two models show a remarkable overlap in performance, hence suggesting that both schemes are equally capable of performing the posed task of score difficulty analysis from sheet music images. In this regard, further work should explore other encoding alternatives to assess whether this performance stagnation is due to the representation capabilities of the considered embedding layers or due to the recognition framework.

\subsection{Multi-task learning experiment}
\label{exp:multi}


In this second study, we assess the capabilities of the multi-task framework proposed in Section~\ref{method:multi} trained simultaneously on the CIPI, PS, and FS datasets for the two \multifc{} and \multicnn{} encoding schemes. Table~\ref{tab:multi} provides the results obtained.

\begin{table}[h!]
\centering
\resizebox{.9\columnwidth}{!}{%
\begin{tabular}{llccc}
\toprule[1pt]
\multicolumn{2}{l}{Encoding}  & Acc$_0$ (\%)       & Acc$_1$ (\%)     & MSE        \\ \hline
\multicolumn{2}{l}{$\text{GPT}^\texttt{multi}_\texttt{FC}$}\\
& CIPI & \textbf{40.3(4.3)} & \textbf{82.0(1.4)} & \textbf{1.3(0.1)} \\
& PS   & 35.9(3.1) & \textbf{78.2(3.4)} & \textbf{1.9(0.2)} \\
& FS   & 45.8(2.5) & 92.0(1.4) & \textbf{0.8(0.1)} \\
\hdashline
\multicolumn{2}{l}{$\text{GPT}^\texttt{multi}_\texttt{CNN}$ }\\
& CIPI & 34.9(5.0)  & 81.4(1.3) & 1.4(0.1) \\
& PS   & \textbf{35.9(2.8)}  & 74.5(3.4) & 2.7(0.2) \\
& FS   & \textbf{45.9(1.2)} & \textbf{92.4(2.1)} & 0.8(0.1) \\
\bottomrule[1pt]
\end{tabular}
}
\caption{Results of multi-task learning experiment when evaluated on different test collections for the two encoding schemes. Bold values highlight the best results per collection and metric.}
\label{tab:multi}
\end{table}

Overall, the \multifc{} method had higher results than the \multicnn{} method on the CIPI and PS datasets, especially on Acc$_0$ and Acc$_1$. For CIPI, \multifc{} surpassed \multicnn{} with gains of 5.4\% in Acc$_0$, 0.6\% in Acc$_1$, and 0.1 in MSE. For PS, \multifc{} slightly exceeded \multicnn{} with a 3.7\% improvement in Acc$_1$ and a 0.6-point reduction in MSE, while Acc$_0$ was nearly equal for both methods, although \multicnn{} had a smaller standard deviation. Both methods displayed similar performance on the FS dataset, with less than a 1\% difference across all metrics. As a result, subsequent experiments will reference the \multifc{} model.

The comparison between Tables~\ref{tab:representation} and~\ref{tab:multi} shows a trend change with better results performed with the FC version of the models.  The other major difference is the relative improvement between the \multifc{} method and the best previous model \cnn{} in the CIPI and slightly in the PS dataset. In contrast, the FS dataset results remain comparable. In CIPI, Acc$_0$ is 11.3\% higher in \multifc{}, and in PS, there is a relative improvement of 12.8\%. For CIPI, Acc$_1$ sees a minor increase of 0.4\%. MSE exhibits a small improvement of 3.6\% for CIPI and 0.5\% for PS.  Possible reasons include label quality differences---CIPI annotated by a musicology team, PS labels provided by the platform, and FS crowdsourced by users---or the impact of dataset sizes---CIPI being the smallest and FS the largest.

\subsection{Ranking generalization experiment}
\label{exp:zero}

In this experiment, we assess the ranking capabilities of the proposal in a zero-shot setting by utilizing the embeddings of the projection layer of the model (check Figure~\ref{fig:down}). We reduce the 768-dimensional embeddings to a single dimension using Principal Component Analysis (PCA) and employ the resulting values to rank the target pieces.

Table~\ref{tab:zero} shows the results obtained resorting to the Kendall rank correlation coefficient, $\tau_c$, for all data collections discussed in the experiment, considering both the single-task and multi-task frameworks posed. Note that MK and HV are only used for benchmarking purposes.

\begin{table}[h!]
\addtolength{\tabcolsep}{-.4em}

\centering
\resizebox{\columnwidth}{!}{%
\begin{tabular}{lccccc}
\toprule[1pt]
                           \multirow{2}{*}{Train} & \multicolumn{5}{c}{Evaluation}\\
                           \cmidrule(r){2-6}  
 & CIPI      & PS        & FS        & MK           & HV             \\ 
\cmidrule{1-6}
\multicolumn{1}{l}{CIPI}  & .67 (.01) & .56 (.02) & .56 (.01) & .67 (.05)    & .50 (.05)    \\
\multicolumn{1}{l}{PS}    & .67 (.01) & .58 (.02) & .56 (.01) & .68 (.01)    & .43 (.04)    \\
\multicolumn{1}{l}{FS}    & .64 (.04) & .55 (.01) & .56 (.02) & \textbf{.71 (.02)}    & \textbf{.56 (.07)}    \\
\hdashline
\multicolumn{1}{l}{MULTI} & \textbf{.68 (.02)} & \textbf{.59 (.02)} & \textbf{.56 (.01)} & .63 (.02)    & .51 (.07)    \\
\bottomrule[1pt]
\end{tabular}
}
\caption{Zero-shot ranking results. Bold values denote the best-performing result on each evaluation dataset.}
\label{tab:zero}
\end{table}

In the three training datasets, the multi-task architecture \multifc{} achieves the best performance with CIPI ($\tau_c=0.68$), PS ($\tau_c=0.59$), and FS ($\tau_c=0.56$). Unexpectedly, the FS method outperforms others in the datasets of the MK ($\tau_c=0.61$) and HV ($\tau_c=0.56$). This outcome may suggest that simultaneous training on all three datasets could limit generalizability. Alternatively, the presence of license-free pieces composed after 1900 in the FS dataset, which users have uploaded, might explain the difference.

The HV dataset displays notably lower generalizability, possibly due to the smaller number of pieces, resulting in higher standard deviations. Potential bias similar to MK could also arise from the predominance of pre-20th-century data in CIPI and PS. These factors might affect the zero-shot experiment's performance. However, we must also acknowledge that most composers used for training are white males, and the HV results are significantly worse than the rest of the datasets. Therefore, future research should investigate and minimize the potential gender gap in difficulty prediction tasks.

\subsection{Comparison with previous approaches}
\label{exp:cipi}


This last experiment compares the goodness of the proposed methodology in sheet music scores against other image-based approaches and with the symbolic-oriented methods domain. 
Regarding sheet image methods, we consider the reference method by Tsai et al.~\cite{ismir2020tsai} based on bootleg mid-representation, denoted as \tj{}. Concerning the symbolic baseline, we reproduce the approach in~\cite{anonym} that proposes to describe the symbolic score in terms of piano fingering information, expressive annotations, and pitch descriptors to feed a recurrent model based on Gated Recurrent Units with attention layers (referred to as $\text{GRU+Att}$). 
Table~\ref{tab:last} provides the results obtained. For comparative purposes, we only consider the CIPI dataset as the reference symbolic work accounted for that collection.


Examining the experiments, the \multifc{} model may be observed to outperform the other cases in the Acc$_0$ figure of merit. However, for the rest of the metrics, the reference symbolic case---denoted as $\text{GRU+Att}$---outperforms all image-oriented recognition models. Such a fact suggests that, while a bootleg score somehow suits this difficulty estimation task, a performance gap between this representation and pure symbolic notation needs to be addressed.

\begin{table}[h!]
\centering
\resizebox{.95\columnwidth}{!}{%
\begin{tabular}{llccc}
\toprule[1pt]
Case  && Acc$_0$ (\%)  & Acc$_1$ (\%)       & MSE        \\ \hline 
\multicolumn{2}{l}{\textit{Symbolic}~\cite{anonym}} \\
& $\text{GRU+Att}$ & 39.5(3.4) & \textbf{87.3(2.2)} & \textbf{1.1(0.2)}  \\
\hdashline
\multicolumn{2}{l}{\textit{Tsat et al.}~\cite{ismir2020tsai}} \\
 & $\text{GPT}_\texttt{EMB}$ & 19.7(4.0) &  58.1(7.2) & 3.3(0.8)\\
 \hdashline
\multicolumn{2}{l}{\textit{Proposal}} \\
& $\text{GPT}_\texttt{FC}$  & 34.3(6.1) & 78.1(4.6) & 1.6(0.3) \\
& $\text{GPT}_\texttt{CNN}$  & 36.2(8.2) & 81.7(1.5)  & 1.4(0.1) \\
& $\text{GPT}^\texttt{multi}_\texttt{FC}$ & \textbf{40.3(4.3)}  & 82.0(1.4) & 1.3(0.1) \\
\bottomrule[1pt]
\end{tabular}
}
\caption{Performance results for the symbolic~\cite{anonym} and Tsai et al.~\cite{ismir2020tsai} methods as well as the proposed approach for the CIPI dataset. Bold values highlight the best result per figure of merit.}
\label{tab:last}
\end{table}

Finally, the \tj{} model achieves the lowest performance of all alternatives, with remarkably lower accuracy rates than our proposal. Note that such a fact emphasizes the relevance of our work as a more suitable approach for performing difficulty estimation in sheet music images.

\section{Conclusions}
\label{sec:conclusion}

Estimating the performance difficulty of a music piece is a crucial need in music education to structure the learning curriculum of the students adequately. This task has recently gathered attention in the Music Information Retrieval field, given the scarce existing research works devoted to symbolic machine-readable scores. However, due to the limited availability of this type of data, there is a need to devise methods capable of addressing this task with image-based sheet music.

Attending to its success in related classification tasks, this work considers the use of a mid-level representation---namely, bootleg score---that encodes the content of a sheet music image with a GPT-based recognition framework for predicting the difficulty of the piece. Instead of directly applying this methodology, we propose using specific embedding mechanisms and multi-task learning to reduce the task complexity and improve its recognition capabilities. The results obtained with five different data collections---three of them specifically compiled for this work---prove the validity of the proposal as it yields recognition rates comparable to those attained in symbolic machine-readable scores.

Further work comprises assessing and proposing alternative representations to the bootleg scores (\textit{e.g.}, solutions based on Optical Music Recognition). Also, we consider that using smaller training sequences using hierarchical attention models or weak labels for varying-length piece fragments may report benefits in the process. Finally, the practical deployment of this proposal in real-world scenarios involving real users may report some additional insights about the validity of the proposal.

\section{Acknowledgment}

We want to thank T.J. Tsai and all his students, especially Daniel Yang, for having conducted the prior research on the bootleg score and, above all, for sharing all their work in the interest of Open Science. We are also grateful to Pedro D'Avila for bringing to our attention the work of Alejandro Cremaschi related to the Hidden Voices project. Lastly, we thank Alejandro Cremaschi and the University of Colorado Boulder Libraries team, David M. Hays and Jessica Quah, for providing us with the scores.

This work is funded by the Spanish Ministerio de Ciencia, Innovación y Universidades (MCIU) and the Agencia Estatal de Investigación (AEI) within the Musical AI Project – PID2019-111403GB-I00/AEI/10.13039/501100011033 and the Basic Science Research Program through the National Research Foundation of Korea (NRF) funded by the Korea Government (MSIT) (NRF-2022R1F1A1074566).

\bibliography{template}

%
%
%

\end{document}